# Mutual Influence of Different Hydrogen Concentration in α-Zirconium System with Vacancies

Heng-Yu Li[12†], Hao-Jun Jia[12†], Yan-Zhen Zhao[12], L.A. Svyatkin[1], I.P. Chernov[1]

[a]Department of General Physics, School of Nuclear Science and Engineering, National Research Tomsk Polytechnic University, Tomsk 634050, Russian Federation

[b]College of Physics, Jilin University, Changchun, 130012, China

*Corresponding author, E-mail: lihengyu.jlu@gmail.com,

[†]These authors contribute equally to this work

## Abstract

Hydrogen embrittlement in zirconium alloys is one of current challenges of nuclear reactor containment and can lead to a rapid decrease in mechanical properties of materials. Although many previous studies investigated the mechanism of hydrogen embrittlement at high hydrogen concentration, few studies concentrate on the mechanism at low hydrogen concentration which is before the formation of hydrides (only in the hydrogen complex state). The purpose of this study is to illuminate the formation mechanism of hydrogen complex. We calculate the formation energy $E_{vac}$, binding energy $E_H$, $E_{H-vac}$ using ABINIT software package. For hydrogen atom at different states in lattice, we obtained $E_{vac}$ increased from 26.3% to 84.6% and $E_{H-vac}$ increased from 40.6% to 119.3%. This indicates that the bond between the original Zr atoms is weakened by hydrogen atoms entering the lattice and formation of vacancy is more possible. In addition, vacancies make hydrogen atoms and Zr atoms more closely connected and further increases the vacancies formation. Therefore, hydrogen complexes tend to expand in lattice, the brittleness of hydrogen complex becomes the main cause of hydrogen embrittlement.

## Introduction

For decades, transition metals are favored to be used in many fields by their excellent properties. However, a challenging problem arising in this domain is these transition metals have fairly strong ability to absorb hydrogen, in some conditions this ability is what we expect but it is also double-edged. Hydrogen absorption capacity can alter the mechanical properties of metals to the point of rendering them unreliable [1], which is so-called hydrogen embrittlement.

Hydrogen embrittlement [2-4] can be classified into several categories by different mechanisms. In our study, we mainly investigate one type of hydrogen embrittlement – embrittlement resulting from forming hydrogen-vacancy complexes in metal lattice [5].

This hydrogen complex's structure is brittle phase, so it's easily becomes fracture origin and leads to brittle fracture [6]. It's notable that hydrogen-vacancy complex is different from hydride, the primary difference between these two substances is the concentration of hydrogen in metallic lattice and whether the generate substance has stoichiometry or not.

α-Zirconium is typical transition metal with strong affinity with hydrogen and widely used in covers of fuel cells in nuclear reactors [7], the elements are exposed to high concentrations of hydrogen and high temperatures can exacerbate this process [8]. Hydrogen and vacancies in lattice lead to various hydrogen-vacancy complexes, obviously these complexes are main factor of hydrogen embrittlement.

Several previous researches have investigated the hydrogen embrittlement effect with one hydrogen atom in Zr lattice [8], hydrogen complex in these systems can be expressed as $Zr_nH$. In our study, we repeated the calculation of $Zr_nH$ systems within 2x2x2 lattice (6.5at%) and 3x3x2 lattice (3at%) including atomic and electron structure, positron life time of hydrogen-vacancy complex, charge density distribution with first-principle approach. Our work makes up some gaps in previous work and present an in depth study of hydrogen embrittlement at low hydrogen concentration.

Our calculations in Zr-H; Zr-vac; Zr-H-vac systems are mainly about four parts: optimization of lattice parameters and relaxation of atomic position, formation energy and binging energy, positron lifetime and charge density distribution.

## Computational methods

All calculations in this work were performed based upon DFT in the generalized gradient approximation (GGA), and first principles calculations were carried out using the ABINIT software package [9], in which Perdew–Burke–Ernzerh (PBE) [10] was employed. The projector augmented wave (PAW) method [11] was used with a cut-off energy of 680 eV. We adopted a Brillouin zone of 3×3×1 with a Gamma (G) centered k-point mesh. Self-Consistent Field (SCF) calculation were performed with a convergence criterion of $10^{-7}$ Hartree in energy. This energy value is the result of a comprehensive consideration of computational efficiency and calculation errors. The relaxation was set to complete since maximal absolute force tolerance below $5*10^{-4}$ Hartree/Bohr (this corresponds to about $2.5*10^{-3}$ eV/Angstrom).

## Results and discussion

The calculation cells of considered Zr systems represented 2x2x2 block of unit cells (Fig.1) and 3x3x2 block of unit cells (Fig.2) with one hydrogen atom and one vacancy, two sizes of supercells represent two different hydrogen concentrations 6.5at % and 3at %. In the calculation cell of Zr-H system contained 16 Zr atoms and one H atom occupied in a tetrahedral T- interstitial or an octahedral O-interstitial position (Fig.1a). In the containing vacancy state (Fig.1b), lattice site 12 was replaced by vacant and there are three available positions for hydrogen atom, each location has different energy characteristics.

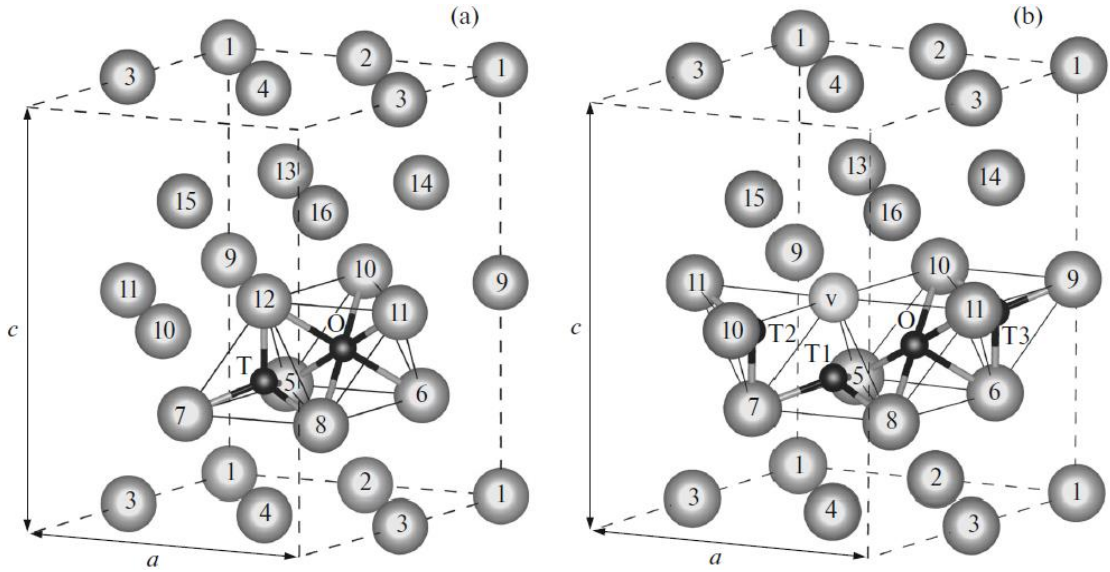

Fig.1 Supercells of Zr-H (a) and Zr-H-Vac system (b), where a and c are supercell's parameters hydrogen atoms in tetrahedral position T (T1 T2 T3) or octahedral position O.

Unitary cell volume for lattice is calculated by equation:
$$V = a^2 \times c \times \cos 30°  \qquad (1)$$
The unitary cell volumes are increased respectively by 0.30% and 0.11%, the dissolution of hydrogen leads to lattice expansion.

Table.1 Transformation rate of lattice parameters and unitary cell volume of Zr

| 16 System | Zr-vac | Zr-H$^O$ | Zr-H$^T$ | Zr-H$^{T1}$-vac | Zr-H$^{T3}$-vac | Zr-H$^O$-vac |
|---|---|---|---|---|---|---|
| lattice parameters | a: -0.34% c: -1.89% | a: 0.03% c: 0.21% | a: 0.27% c: -0.56% | a: 0.15% c: -2.24% | a: -0.09% c: -2.16% | a: -0.37% c: -1.54% |
| unitary cell volume | -2.53% | 0.30% | 1.11% | -1.93% | -2.16% | -2.14% |
| 36 System | Zr-vac | Zr-H$^O$ | Zr-H$^T$ | Zr-H$^{T1}$-vac | Zr-H$^{T3}$-vac | Zr-H$^O$-vac |
| lattice parameters | a: -0.40% c: -0.32% | a: -0.03% c: 0.21% | a: 0% c: 0.38% | a: -0.24% c: -0.36% | a: -0.18% c: -0.56% | a: -0.52% c: -1.26% |
| unitary cell volume | -1.14% | 0.10% | 0.38% | -0.85% | -0.94% | 0.97% |

As for pure Zr system with vacancy, the vacancy formation caused a reduction of lattice, lattice parameters a and c decreased by 0.34% and 0.18%, unitary cell volume decreased by 0.25%. The hydrogen dissolution can lead to two opposite changes expansion and reduction depending on the position hydrogen occupied. The rate of transformation is given in the table.1.

Compared data from 16 atoms and 36 atoms we can discover that their variation tendencies are roughly same and the stability of larger lattice is better than the smaller. But we can see the lattice parameters of pure 16 and 36 Zr system have 0.05% bias, this bias is caused by using different k-point during calculation.

These data demonstrate hydrogen dissolution and vacancy formation can cause non-ignorable impact on pure Zr lattice, previous work [8] also proved that there is a large shift of atomic positions in the supercell. The results show that it's necessary to relax the lattice and in subsequent calculations lattice parameters and atomic position data from relaxation will be used.

Formation energy and binging energy are significant standard to measure whether the system is stable. In $Metal_N Vac_n$ systems the formation energy of $Vac_n$ is given by formula:

$$E_{vac} = E(Metal_{N-n}) - \frac{N-n}{N} E(Metal_N) \tag{1}$$

Within $Metal_N H Vac_n$ systems the formation energy of $Vac_n$ is given by formula:

$$E_{vac} = E(Metal_{N-n}H) + \frac{n}{N} E(Metal_N) - E(Metal_N H) \tag{2}$$

The hydrogen binding energy $E_H$ formed by formula:

$$E_H = E(Metal_{N-n}) + \frac{1}{2} E(H_2) - E(Metal_{N-n}H) \tag{3}$$

The hydrogen-vacancy binding energy $E_{H-v}$ formed by formula:

$$E_{H-v} = E(Metal_N H) + E(Metal_{N-n}) - E(Metal_N) - E(Metal_{N-n}H) \tag{4}$$

Where $E(Metal_N)$, $E(Metal_N H)$, $E(Metal_{N-n})$, $E(Metal_{N-n}H)$ are the total energies of Metal, Metal-Vac, Metal-H, Metal-H-Vac system respectively. N and n are the number of lattice sites in supercell we calculated and the number of vacancies.

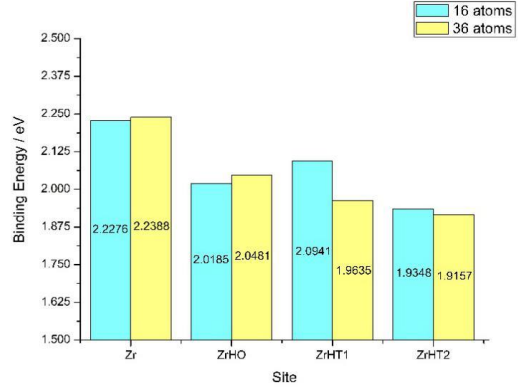

Fig.2 Energy of vacancy formation $E_{vac}$ [eV] in Zr and in the Zr–H system

The formation energy of vacancy is presented in Fig.2, the value of $E_{vac}$ in 16 atoms system agrees well with other's calculation and experiment value, the 36 atoms system there is no data published yet. Primarily through the analysis of 16 atoms system the presence of hydrogen atom in Zr lattice reduced the formation energy of vacancy by 0.13-0.29 eV approximately 5.9%-13.1%, this demonstrates that the presence of hydrogen atoms significantly weakens the bonds between Zr atoms in the lattice, making the formation of vacancies more possible. The vacancy is the main factor of point defect and the hydrogen embrittlement phenomenon appears further. Comparing with 16 atoms the 36 atoms system also produced similar result with 0.19-0.32 eV and 8%-14% reduction of vacancy formation energy, however, the reduction is slightly greater than 16 atoms system.

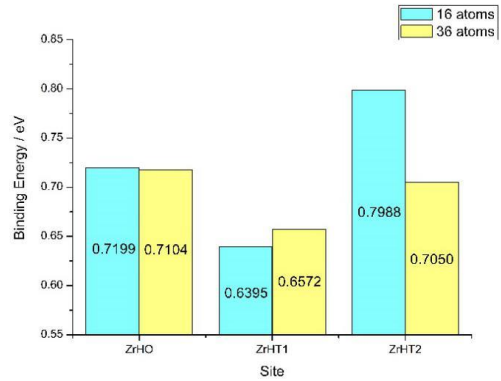

Fig.3 Binding energy of hydrogen $E_H$ [eV]

Fig.3 presents binding energy of hydrogen $E_H$, notably that both in 16 and 36 atoms system the binding energy in system with vacancy is noticeable higher than system without vacancy. In Zr-$H^O$ system $E_H$ increased by 0.21 eV – 40.9% and 0.19 eV – 36.2%, respectively in 16, 32 atoms systems. In 16 atoms Zr-$H^T$ system $E_H$ increased by 0.13 eV – 26.3% at T1 site and 0.29 eV – 57.8% at T2 site. Relative to 16 atoms system the percentage increase of binding energy of 32 atoms system at tetrahedral T site is much greater, $E_H$ of Zr-$H^{T1}$-vac increased by 0.27 eV – 72.1% and the highest percentage increase appears in $E_H$ of Zr-$H^{T2}$ by 0.32 eV – 84.6%. Since binding energy of energy represents the minimum energy is required to disassemble a system of particles into separate parts, the higher binding energy of hydrogen demonstrates that the interaction between H and Zr is stronger. Obviously Zr-$H^{T2}$ is the strongest position H occupies, further analysis combines of H impact on difficulty of vacancy formation: H causes vacancy to form more easily and the formation of vacancy will make H more stable in the lattice in turn, this will lead to the aggregation and expansion of point defects eventually lead to hydrogen embrittlement.

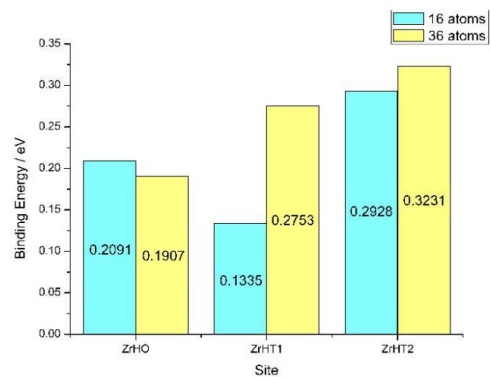

Fig.4 binding energy of Hydrogen–vacancy $E_{H\text{-vac}}$ [eV]

Fig.4 presents binding energy of H-vacancy $E_{H\text{-vac}}$, this table mainly applied to compare the strength of the interaction between vacancy and H. Whether in 16 atoms system or 36 atoms system, the maximum value of binding energy corresponds to the H position, at the T2 tetrahedral site greater by 40.0% and 119.3%, respectively, than at O and T1 sites in 16 atoms system. Vertical contract 16 and 36 atoms system, O and T2 site binding energy of 36 atoms system is similar with 16 atoms system, T1 site greater by 106.2% however also lower than T2 site. This demonstrates that T2 tetrahedral site is the most stable position that the H will occupy and at low hydrogen concentrations T1 tetrahedral site will be more stable than at high hydrogen concentrations, at low hydrogen concentrations point defect it's more likely to exist at T2 site and at high hydrogen concentrations seems to be both T1 and T2 site.

Charge density distribution is a visible and effective method to investigate interaction of hydrogen and vacancy and explain the mechanism of hydrogen embrittlement. In this work, charge density distribution divided into two parts: electron density distribution and positron density distribution. Where electron density distribution is presented by density gradient plots drawn along different lattice planes. For more accurate analyzation of interaction among hydrogen atom vacancy and Zr, we plotted electron density distribution in two lattice planes: (0001) (11$\bar{2}$0). Fig.1-3 present electron density distribution for pure metal, metal with vacancies, as well as for the Zr–H and Zr–H–v systems.

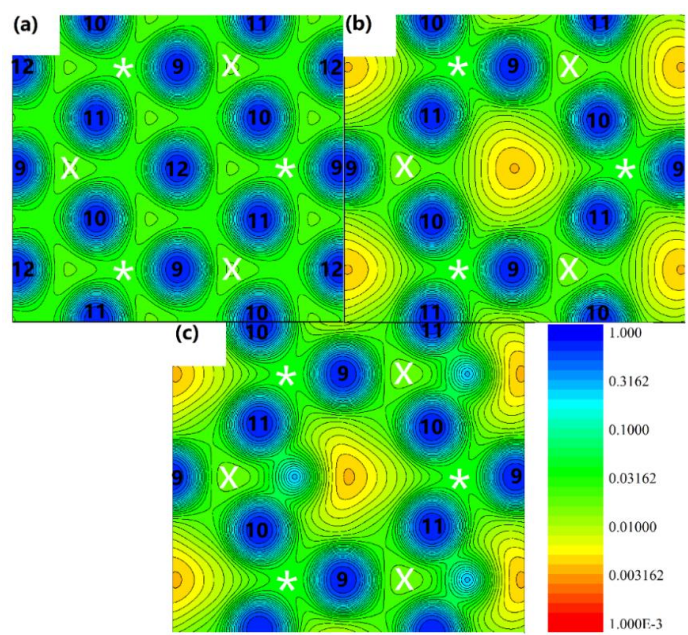

Fig.5 Electron density distribution for pure Zr (a) and for the Zr–v (b) and Zr–H$^{T2}$–v (c) systems in the (0001) plane passing through vacancies and zirconium atoms in 16 atoms systems. Color gradation scale is given in electrons/Bohr$^3$ units.

Compared with pure Zr in Fig.5a the Fig.5b shows that the formation of vacancy in Zr lattice leads to a considerable redistribution of the metal electron density, especially in FCC-sites (X) and HCP-sites (*), these redistributions demonstrate that the interaction between the above-indicated atoms of Zr in the region of FCC-sites becomes weak and their interaction in the region of HCP-sites reinforced.

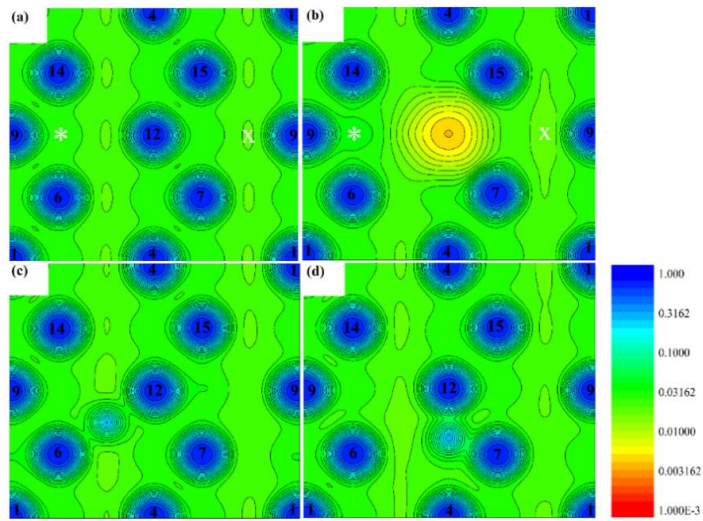

Fig.6 Electron density distribution for pure Zr (a) and for the Zr–v (b), Zr–H$^O$(c), and Zr–H$^T$ (d) systems in the (11$\bar{2}$0) plane passing through vacancies and atoms of zirconium and hydrogen in 16 atoms system. Color gradation scale is given in electrons/Bohr3 units.

In Fig.6 can be clearly seen that hydrogen causes a considerable redistribution of the electron density of the metal. This Fig demonstrates that one can see a higher level of electron density between H atoms and Zr atoms nearest to H: these atoms are enclosed by common isolines of the electron density distribution. This is evidence of a formation of the metal–hydrogen bond whose considerable part is attributed to the covalent component. In the tetrahedral interstitial site, a hydrogen atom is bound stronger with zirconium atoms than in the octahedral interstitial site, since in the first case, the atoms of hydrogen and zirconium are enclosed by a greater number of common isolines. It correlates with the results of calculations of the hydrogen–zirconium binding energy: $E_H(T) > E_H(O)$.

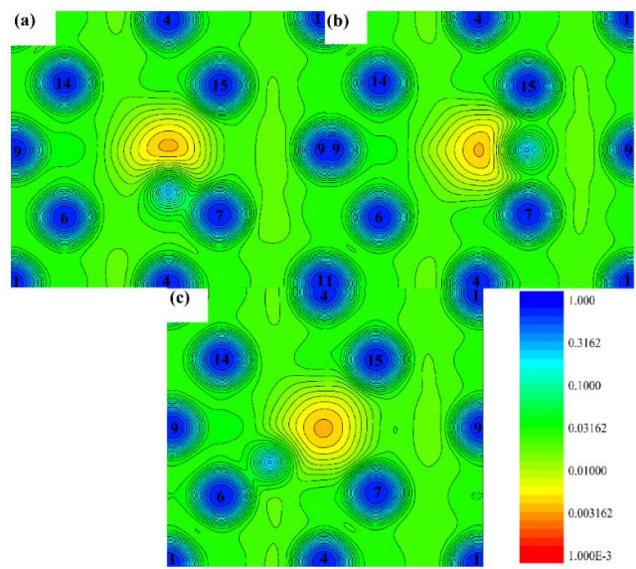

Fig.7. Electron density distribution for the Zr–H$^{T1}$–v (a), Zr–H$^{T2}$–v (b), and Zr–H$^O$–v (c) in the (11$\bar{2}$0) plane passing through vacancies, atoms of zirconium and hydrogen in 16 atoms system. Color gradation scale is given in electrons/Bohr3 units.

The vacancy formation leads to the break of the common contour of isolines enclosing pairs of zirconium atoms 5–13, 7–15, and 8–16, situated above and below it, which means that the bond between these atoms becomes weaker. The presence of hydrogen at the T2 tetrahedral interstitial site near a vacancy due to the lattice relaxation reconstructs this contour for a pair of zirconium atoms 15 and 7, situated above and below the hydrogen atom (Fig.7b), which strengthens the bond between these atoms. In the case of the Zr–$H^{T1}$–v systems (Fig.7a) and Zr–$H^{O}$–v (Fig.7c), this contour remains broken. Apparently, it explains why the hydrogen-zirconium energy binding is maximal when hydrogen is arranged at the T2 tetrahedral interstitial site.

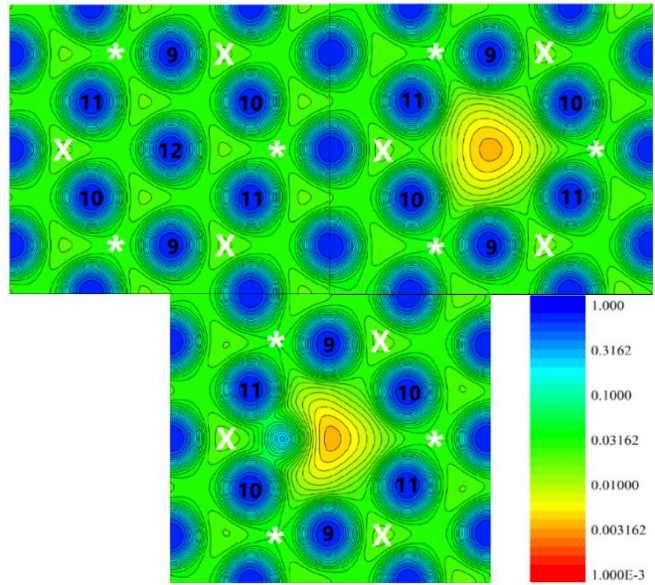

Fig.8 Electron density distribution for pure Zr (a) and for the Zr–v (b) and Zr–$H^{T2}$–v (c) systems in the (0001) plane passing through vacancies and zirconium atoms in 36 atoms systems. Color gradation scale is given in electrons/Bohr$^{-3}$ units.

Fig.8 we present the electron density distribution, from Fig.4 we can find that although compared with 16 atoms systems (Fig.5) the 36 atoms systems' unit cell is larger, the electron density distribution is similar. It indicates that whether at 3at% or 6at% hydrogen concentration the effect of hydrogen on the Zr lattice is localized in the cell which hydrogen atom exists in. And it is worth mentioning that in 36 atoms systems the electron density more concentrated on HCP site (*), the electron density in region of vacancy is lower, and in Zr–$H^{T2}$–v (c) system the isoline near hydrogen atom is more intensive, these distinctions of electron density distribution suggest that the 36 atoms system at 3at% hydrogen concentration is more likely to form defect.

## Conclusion

We plotted electron and positron density distribution in order to analyze the influence of hydrogen atoms and vacancies on α-Zr. It established that the presence of vacancies and hydrogen atoms deviate the atoms of α-Zr from original positions. In addition, the presence of hydrogen atoms reduces the formation energy of vacancies while the presence of vacancies increases the binding energy of hydrogen atoms. The results demonstrate that hydrogen atoms can weaken the bonds among Zr atoms in Zr systems, which leads to the formation of vacancies more possible and the formation of vacancies makes hydrogen atoms more stable in Zr systems. Hydrogen-vacancy complexes lead to the aggregation and expansion of point defects, eventually leading to hydrogen embrittlement. Although different hydrogen concentrations have similar formation process in hydrogen-vacancy complex, there are still slightly different effects that hydrogen-vacancy complex more likely to form at 3at%, especially in T2 site. Electron density distribution establishes that the presence of vacancies strengthens the covalent part of bond between the hydrogen atoms and zirconium atoms, and the presence of hydrogen atoms reduces the electron density. Moreover, at 3at% hydrogen concentration, these two kinds of effects are more noticeable.

## Conflicts of interest

There are no conflicts of interest to declare.

## Acknowledgment

The research was funded by the Tomsk Polytechnic University Competitiveness Enhancement Program.